\documentclass[referee,a4paper,12pt,traditabstract]{swsc} 

\usepackage{graphicx}
\usepackage{txfonts}
\usepackage{subfigure}
\usepackage{epstopdf}
\usepackage{lineno}
\usepackage[authoryear,round]{natbib}
\usepackage[backref]{hyperref}
\usepackage{url}

\hypersetup{colorlinks=true,citecolor=cyan,urlcolor=cyan,linkcolor=blue}

\begin{document}


   \title{Solar Spectral Irradiance Variability in Cycle 24: Observations and Models}
   
   \titlerunning{Solar Variability in Cycle 24}

   \authorrunning{Marchenko et al.}

   \author{S. V. Marchenko\inst{1,2}\fnmsep\thanks{Corresponding author},  M. T. DeLand\inst{1,2}  \and J. L. Lean\inst{3}
          }

   \institute{
          Science Systems and Applications, Inc., Lanham, Maryland, USA\\
          \email{\href{mailto:sergey\_marchenko@ssaihq.com,matthew.deland@ssaihq.com}{sergey\_marchenko@ssaihq.com,matthew.deland@ssaihq.com}}
          \and
          NASA Goddard Space Flight Center, Greenbelt, Maryland, USA\\
          \and
           Space Science Division, Naval Research Laboratory, Washington, District of Columbia, USA\\
           \email{\href{mailto:judith.lean@nrl.navy.mil}{judith.lean@nrl.navy.mil}}
}


  \abstract
{
Utilizing the excellent stability of the Ozone Monitoring Instrument (OMI), we characterize both
short-term (solar rotation) and long-term (solar cycle) changes of the
solar spectral irradiance (SSI) between 265-500 nm during the on-going
Cycle 24. We supplement the OMI data with concurrent observations from the GOME-2 and 
SORCE instruments and find fair-to-excellent, depending on wavelength, agreement 
among the observations and predictions of the NRLSSI2 and SATIRE-S models.}


   \maketitle

\section{Introduction}
The egress from solar cycle 23 to the current Cycle 24 has been discussed extensively in the literature, especially in terms of the unexpected wavelength- and time- dependence of the ensuing SSI changes observed by the Solar Radiation and Climate Experiment (SORCE) \citep{{harder2009},{lean2012},{erm2013},{sol2013},{mor2014},{yeo2014},{woods2015}}. Recent studies \citep{erm2013} note the importance of the $\lambda220-400$ nm range for climate modeling, as well as the
lack of extended, regular and sufficiently accurate data records in the $\lambda>$400 nm domain \citep{yeo2015} and references therein). Here we attempt to partially remedy the situation by analyzing daily solar observations made in Cycle 24 by relatively `untapped' sources, OMI and GOME-2, comparing them with both the SORCE observations and model predictions in the 115-1500 nm spectral range which is a critically important input for model simulations of climate change. This wavelength region accounts for $\sim$95\% of the total solar irradiance, thus spanning a range of terrestrial influence from chemistry of the upper atmosphere down to lower atmosphere heating and surface response \citep{erm2013}. Moreover, the contribution of variability in the relatively narrow 150-400 nm region to long-term total solar irradiance changes remains poorly known, in the range 30\% \citep{lean1997a} to 60\% \citep{mor2011}. 

\section{Models and Observations}
In the following study of the SSI variability we focus on the on-going solar cycle 24, utilizing daily solar observations provided by three space missions (OMI, GOME-2 and SORCE), and relating them to the outputs from two models, NRLSSI2 and SATIRE-S. For a broader perspective we also analyze SSI changes during the prior solar cycle 23, to assess the robustness of our results in cycle 24.

\subsection{The Models}
Among the numerous models \citep{erm2013} developed to specify the SSI variability, we select two as representatives of the  two broad classes \citep{{yeo2014},{woods2015}}: the purely empirical, proxy-based  (NRLSSI2), and semi-empirical, more physics-based (SATIRE-S). Both models, the original NRLSSI \citep{{lean1997},{lean2000}}, then NRLSSI2 \citep{cod2016} and SATIRE-S \citep{{unruh1999},{yeo2015}} reconstruct SSI changes by estimating the alteration of the net (disk-integrated) solar emission at a given wavelength that arises from bright features (faculae) and dark features (sunspots) present at any given time on the Sun$^\prime$s disk. However, the two models estimate the time- and wavelength-dependent contributions from facular brightening and sunspot darkening in different ways. In the NRLSSI2 model these components are derived from linear regression models that connect sunspot and facular proxy indices to observed SSI modulation by solar rotation (thereby avoiding spurious relationships from instrumental trends in the observations); the rotationally-modulated relationships are then scaled to the solar cycle, self consistently with independent models of total solar irradiance variations, which faculae and sunspot similarly modulate: see \citet{cod2016} for more details. SATIRE-S utilizes physical models of the solar atmosphere to estimate the intensity spectra for the quiet Sun, and for the faculae (network inclusive) and sunspot features, additionally distinguishing between the sunspot$^\prime$s umbral and penumbral regions. The approach for deriving the corresponding surface coverage of these features is to identify sunspot features in continuum intensity images and identify faculae as regions in magnetograms where there are no spots, and where the magnetic flux exceeds a specified threshold. Solar irradiance is then determined from disk integration of emissions from all features. Both models apply additional intensity offsets and scaling in order that the reconstructed spectra match the absolute levels of observed reference spectra.

\subsection{Observations: OMI}
The Ozone Monitoring Instrument \citep{lev06} has been operating on the Aura remote-sensing satellite since July 2004, 
acquiring mid-resolution, $\delta\lambda\sim$0.4-0.6 nm, backscattered Earth radiance spectra in the 264-504 nm range. For calibration purposes the Sun is observed once per day. The broad field of view (FOV) provides 30-60 simultaneously recorded, disk-integrated solar spectra in 3 instrument channels \citep{dob2006}: 264-311 nm (hereafter mentioned as UV1), 307-283 nm (UV2) and 349-504 nm (VIS). These channels were radiometricly calibrated using a specifically constructed high-resolution solar spectrum \citep{dob2008b} and have absolute accuracies better than 4\% in the 270-500 nm range \citep{dob2008a}. Transitions between adjacent spectral channels are constrained to be smooth to $^<_\sim$0.5\%.  Comprehensive, multi-parametric tracking of the long-term performance \citep{schen2016} demonstrates high instrument stability and low optical degradation, $\sim$0.2-0.5\% yr$^{-1}$ for the data used in this study. 

Here we use observations from this well-characterized instrument to provide SSI data for Cycle 24. We note that the quoted degradation rates of 0.2-0.5\% yr$^{-1}$, attest, among other metrics \citep{schen2016}, to high (for a hyperspectral space-borne mission) OMI stability. Nevertheless, these gradual, quite predictable changes still exceed the expected solar cycle variability of order 0.1\% (during the five years from cycle minimum to maximum) in solar spectral irradiance at $\lambda >$300 nm. Even after degradation corrections \footnote{The daily degradation-corrected OMI data are available at: http://sbuv2.gsfc.nasa.gov/solar/omi/}, the OMI irradiances show systematic biases of as much as $\sim 0.2\%$ on solar-cycle time scales. It is for this reason, in addition to accounting for instrument degradation, that our analysis also assesses SSI changes associated with solar rotation, since over these (much) shorter time scales instrumental effects are minimal.  

\subsection{GOME-2}
The Global Ozone Monitoring Experiment-2 (GOME-2: 240-790 nm spectral range, with $\delta\lambda\sim$0.3-0.5 nm resolution) is a part of the Metop series of the remote-sensing satellites \citep{munro2015}. 
We select for analysis the daily solar (not adjusted for degradation) observations provided by GOME-2 on Metop-A (launched in October 2006).   

\subsection{SORCE}
We supplement the OMI and GOME-2 spectra with observations made by the Solar Radiation and Climate Experiment (SORCE: launched in January, 2003). In particular, we use the wavelength-binned, degradation-corrected data from the Solar Stellar Irradiance Comparison Experiment (SOLSTICE; 115-310 nm spectral range, $\delta\lambda=$0.1 nm; the publicly available product is binned to 1 nm) \citep{{mccl2005},{snow2005}}, and the Solar Irradiance Monitor (SIM; 240-2400 nm spectral range, $\delta\lambda=$0.6-24.6 nm) \citep{harder2005}.

\subsection{Assessment of the Long-term (Solar Cycle) SSI Variability in Cycle 24}
Evaluating the solar cycle (long-term) SSI variability from observations, we exclusively rely on the OMI data. The relatively higher (by almost an order of magnitude compared to OMI) GOME-2 degradation rates \citep{munro2015} 
impairs assessment of the long-term SSI variability with the required $<<$0.5\% accuracy. Numerous studies \citep{{erm2013},{yeo2014},{yeo2015},{woods2015}} show some unresolved instrumental problems affecting the long-term SORCE/SIM and SORCE/SOLSTICE records.

\begin{figure}
   \centering
    \includegraphics[width=1.0\columnwidth]{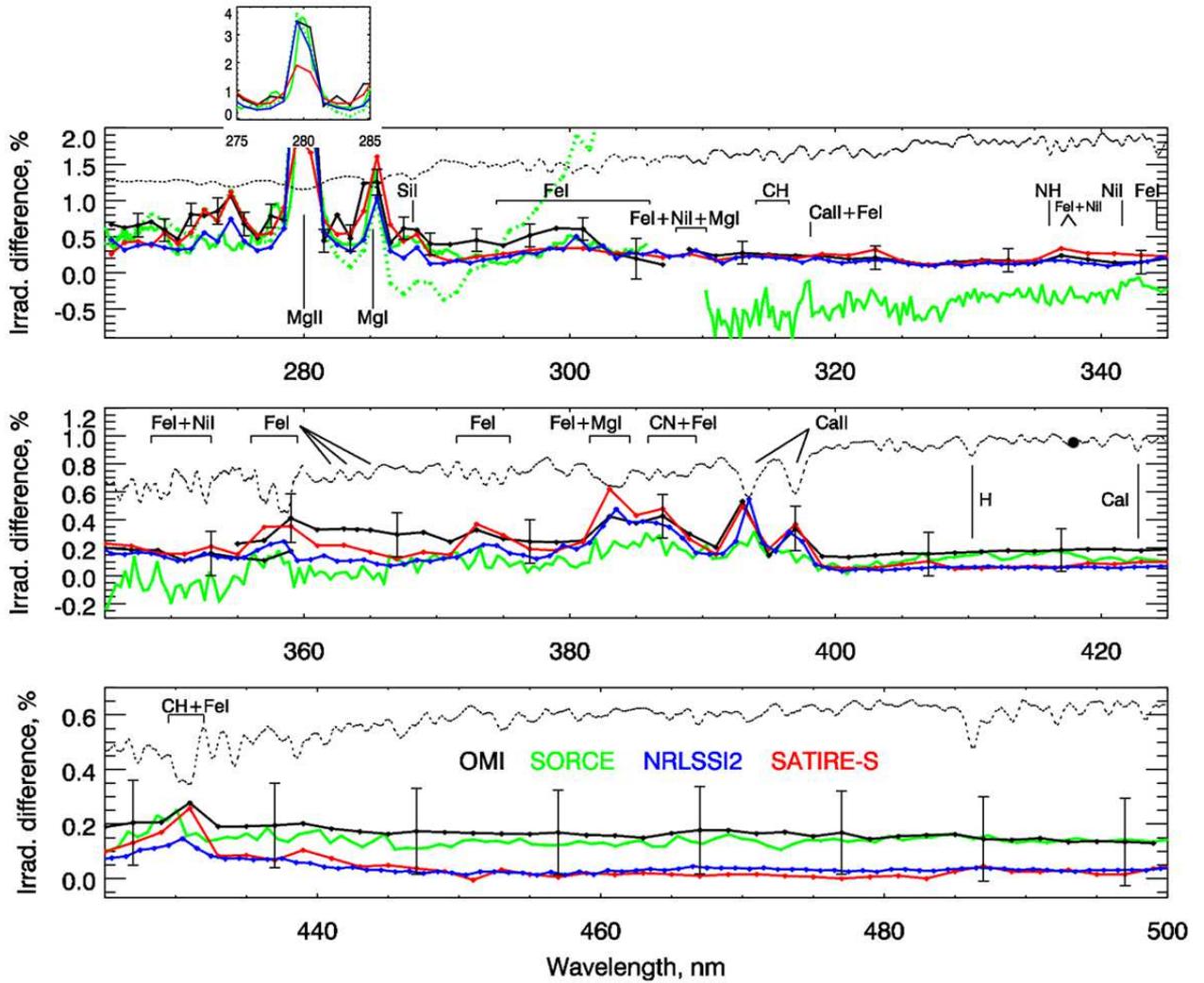}
   \caption{\small 
The averaged and normalized long-term (yy2012-2014 vs. yy2007-2009) SSI changes, as observed by OMI (black lines with representative $\pm 1\sigma$ error bars) 
 and predicted by the models. For reference, the  scaled OMI solar spectrum is shown as a dotted line, along with the marked major solar lines and line blends. The small insert shows the full-scale variation of the Mg II line doublet. Dotted green lines follow the SORCE SOLSTICE (v15) data; full green lines show the SORCE SIM (v22) set.} 
   \label{fig:fig1}
   \end{figure}

To quantitatively assess long-term SSI variability, we merge the OMI solar observations in a single daily record applying the method described in \citet{mar2014}. We follow \citet{dob2008a} in choosing the borders of the spectral channels, hence avoiding production of duplicate values in the regions of spectral overlaps. However, we use the data in the overlapping areas to access uncertainties introduced by the degradation model. In each spectral channel we apply a wavelength- and FOV-dependent (recall that OMI simultaneously registers 30-60 FOVs) degradation model \citep{mar2014}, then bin the corrected irradiances at pre-determined wavelengths. Locations and widths of the chosen spectral bins conform to outputs of the solar models. Each individual (30-60 FOVs) solar spectrum is sampled at a slightly different ($\sim$1/10 of the spectral element width) wavelength grid, thus providing an opportunity to screen the binned data for $\pm 1.5\sigma$ outliers. After applying the optical degradation model we make no further attempts to adjust the differences between the OMI channels. One may notice that in Figure 1 there are no apparent biases exceeding the typical $\pm 1\sigma$ uncertainties in the first transition region around 310 nm. However, there is a distinctive $\sim$0.15\% step around 350-360 nm in the second transition. Assessing various contributing factors, we find that systematic channel-to-channel differences almost always dominate other error sources. This provides an upper 0.2\% limit for errors in the degradation-corrected OMI irradiances \citep{mar2014}, an improvement over the typical degradation-related uncertainty estimates for various space missions:  e.g., 1-2\% for UARS/SUSIM \citep{{flo1998},{mor2011}}; see also the compilation (Table 3) in \citet{yeo2015}.

In developing a degradation model for the OMI observations, we observe that throughout the mission both OMI irradiances and radiances consistently follow nearly linear long-term trends \citep{schen2016}, to within the uncertainties mostly dictated by seasonal fluctuations (either the goniometric changes in solar observations or some geophysical factors). We therefore assume that during the prolonged solar yy2007-2009 minimum (when solar irradiance variations are minimal) all the long-term ($\sim$2 years) changes in irradiances can be ascribed to instrument sensitivity changes and represented by a linear function, which is then extrapolated forward in time. Additional, independent observations support this basic assumption; many hyperspectral instruments acquiring near-UV and visible spectra show optical degradation changes that can be readily approximated by linear trends at the late stages of long-lasting ($\gg$5 years) space missions. 

The daily OMI solar observations were made without interruption between July 2006 and March 2016. We find \citep{schen2016} that, analogous to the SUSIM case \citep{{flo1999},{kriv2006}} the total solar exposure time governs the throughput changes in the optical pathway assigned to acquisition of the solar data. Over the mission time these changes amount to 3-7\% (VIS-to-UV), or only half that, since in this study we extrapolate the degradation model exclusively on the cycle-24 epoch. Thus, for the considered $\sim$5 year time span the quoted 0.2\% uncertainties effectively absorb the relatively smaller, $^<_\sim$0.1\% systematic errors stemming from linear approximation of the degradation trends. 

After adjusting the daily solar records for instrument degradation, we group them into monthly averages, excluding from consideration the data taken around the months of November and December so as to avoid periods of extreme solar incident angles on the OMI diffusors. We apply exactly the same time- and wavelength binning approaches to observations and the relevant model output. We use the solar-minimum epoch between July 2007 and September 2009 as a reference period against which to quantify SSI changes at other times during the solar cycle. Starting from July 2007, we produce monthly SSI averages, then, month-by-month, subtract the SSI references from the data. Hence, for each month we produce the following normalized differences:  
d1=(July2013 - July2007)/July2007,  d2=(July2013 - July2008)/July2008, d3=(July2013 - July2009)/July2009, and so on. For each given month we average d1, d2 and d3. The described procedure provides normalized monthly-mean differences, referenced to the solar minimum between the cycles 23 and 24. We further group these differences for the epoch January 2012 - October 2014 and plot the average in Figure 1. The representative uncertainties shown at each OMI channel in Figure 1 indicate individual errors from each wavelength bin additionally adjusted for the 0.15\% inter-channel biases. We use the same routine on the SORCE data and show them in Figure 1. Consistent with the findings of \citet{{yeo2014b},{yeo2015}} and \citet{woods2015}, we note that in most cases the cycle-24 SSI variability derived from the SORCE data cannot be reconciled with model predictions shown in Figures 1 and 2. Nor can we reconcile SORCE$^\prime$s SSI variability with the solar-cycle SSI changes detected in the OMI observations, particularly in the $\lambda$300-350 nm range.    

In the next step of the analysis, we select from the monthly max-mean differences those spectral regions occupied by strong spectral blends ($\lambda\lambda =280,285,393$ nm), as well as relatively line-free regions ($\lambda\lambda =267.5,313,340,442,471,500$ nm) and compare the wavelength-binned, time-depended variations in Figure 2, where we denote the strong spectral transitions with `l' and the relatively line-free regions with `c'. We supplement these narrow-band measurements with the broadband data (the low panels of Figure 2). In Figures 2 and 3 we also show 1$\sigma$ errors for some representative wavelength bins, derived as a standard deviation of data contributing to the given bin.    

  \begin{figure}
   \centering
    \includegraphics[width=1.0\columnwidth]{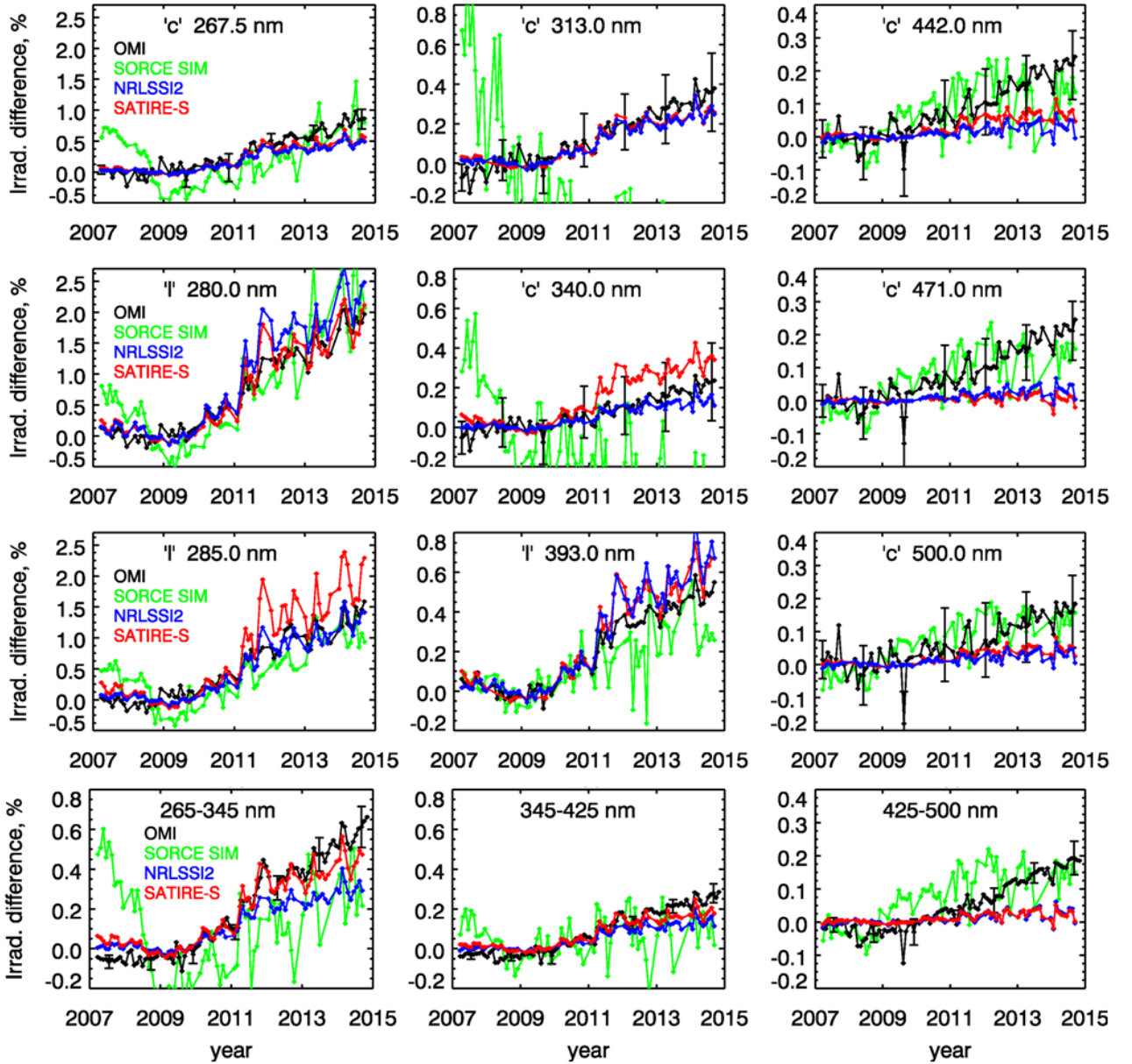}
   \caption{\small The normalized, wavelength- and time-binned OMI fluxes (with representative $\pm 1\sigma$ error bars) compared to the consistently sampled model outputs and the SORCE SIM data. The strong spectral transitions are denoted with `l' and the relatively line-free regions with `c'.} 
   \label{fig:fig2}
   \end{figure}

\subsection{Assessment of the Short-Term (Rotational) SSI Variability in Cycles 23 and 24}

For both the models and observations (OMI, GOME-2 and SORCE), the short-term SSI variability is assessed by following the approach of \citet{mar2014}, thereby effectively minimizing the effects of instrument degradation on the inferred SSI changes. We choose 8 well-defined rotational cycles between July 2012 and April 2013. All the chosen cycles [marked in Figure 1 from \citet{mar2014}] show similar amplitudes of the rotational SSI modulation comparable (by a factor of two) to the amplitude of the long-term SSI changes in cycle 24. For each rotational cycle we average the daily solar observations centered on the corresponding maximum and the two adjacent minima (from 2 to 10 days for a given maximum or minimum). Thus, for each maximum we produce two normalized differences, (max-min1)/min1 and (max-min2)/min2, then average them and show the result in Figure 3 along with the accordingly sampled (precisely matched dates) and averaged model outputs. 

Though the original OMI and GOME-2 data offer $\sim$0.5 nm spectral resolution, we bin the observed SSI changes to the 1-2 nm wavelength steps, thus reproducing the model wavelength grids. When inspecting the observed rotational SSI variability patterns in Figure 3, it is necessary to consider the large differences between the higher spectral resolution observations that produce the binned OMI and GOME-2 SSIs and the much lower, wavelength-dependent resolution of the SORCE data, specifically in the $\lambda > $300 nm domain. For example, compared to SORCE data, the OMI and GOME-2 data consistently reveal much higher contrast between the strong spectral blend at $\lambda =$430 nm and the adjacent line-free regions. The same applies to H and K CaII lines around $\lambda \sim$395 nm (Figure 3). As a further example of spectral resolution impact, we note that the substantially higher-resolution solar spectra show solar-cycle variability in excess of 2\% in a cluster of solar lines around 540 nm \citep{dan2016}. Once diluted in the low-resolution data, such relatively weak signal drops below noise level in the binned GOME-2 data in Figure 3.  In general, in the $\lambda\lambda $450-800 nm range the imposed 1-2 nm spectral binning systematically diminishes both short-term (Figure 3) and long-term (Figure 1) SSI variability to $<0.2\%$ levels. Moreover, in the high spectral resolution, Sun-as-a-star observations reported by \citet{liv2007}, the solar-cycle modulation at the center of he CaII K line exceeds 30\% (cf. $\sim$0.5\% in the 2-nm binned OMI data in Figure 1).         

  \begin{figure}
   \centering
    \includegraphics[width=1.0\columnwidth]{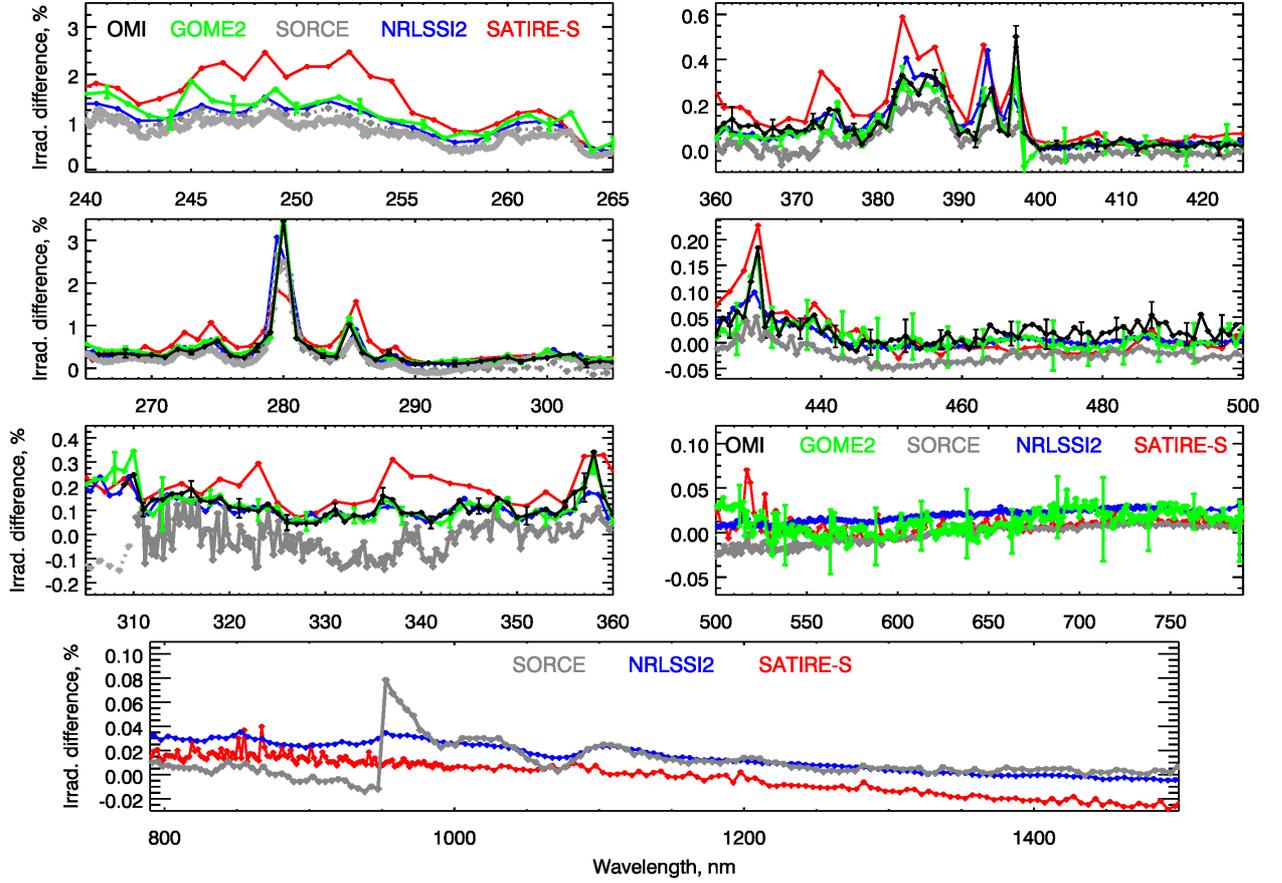}
   \caption{\small The normalized SSI variability spectra compiled from the rotational modulation cycles  for the OMI, GOME-2  and SORCE are shown together with the consistently sampled and adjusted model outputs.
The shown data are averages of 8 rotational cycles selected between June 2012 and April 2013, with representative $\pm 1\sigma$ error bars drawn for the OMI and GOME-2 observations. Dotted grey lines follow the SORCE SOLSTICE (v15) data; full grey lines show the SORCE SIM (v22) set.}
   \label{fig:fig3}
   \end{figure}  

To examine whether the differences between the rotationally-modulated SSI changes observed by OMI and modeled by NRLSSI2 and SATIRE, evident in Figure 3, are consistent in other epochs and observations, we show in Figures 4 and 5 comparisons of the rotational modulation of eight selected SSI wavelength bands measured by instruments on OMI, GOME-2 and SORCE with the NRLSSI2 and SATIRE models. One could argue that these two approaches to evaluation of short-term SSI variability are complementary. The method used in Figure 3 is idealized in a certain way, and focuses on amplitude and spectral dependence by selecting only maximum and minimum values for each rotational modulation.  It does not provide any means for evaluation of differences between individual rotations.  The approach shown in Figures 4 and 5 examines time-resolved SSI variability, where it is evident that rotational modulation features grow and decay over a few months, and that each rotation can have its own temporal structure (e.g. compare the first 3 peaks in Figure 4). Hence, we utilize both the wavelength-resolved but time-binned (Figure 3) and the time-resolved but wavelength-binned (Figures 4 and 5) approaches.

Greatly expanding the relatively limited wavelength ranges in Figures 1-3, we include SORCE data sampling the prominent Lyman$-\alpha$ line (121-122 nm) in Figures 4 and 5. In the SORCE spectral range this particular line shows the highest sensitivity to all phenomena related to development of active solar regions. On the other hand, the 790-1500 nm wavelength bin is dominated by the flux modulations produced by large groups of solar spots, thus serving as an additional test of model predictions along with the shorter-wavelength UV regions responsive to the facula-related phenomena. In addition to providing the obvious benefits of multi-epoch coverage and greatly extended wavelength range, Figures 4 and 5 also test the models against data processed with a completely different algorithm that practically nullifies the biases caused by long-term instrument degradation: each of the observed and modeled time series in Figures 4 and 5 has been detrended by removing 81-day running means. In this way long-term changes arising from both solar cycle activity and instrumental drifts are largely removed, thereby isolating the SSI changes imposed by the Sun$^\prime$s rotation.

\section{Results and Discussion}

\subsection{Long-Term (Solar Cycle) SSI Variability: Observations and Models}

In specifying solar spectral irradiance variability,  both the empirical (NRLSSI2) and semi-empirical (SATIRE-S)  models  employ a system of inevitable  simplifications and assumptions. This approach is necessary because any purely theoretical, physics-based simulations of the time- and wavelength-resolved SSI variability with the fidelity needed for terrestrial applications are still beyond current capabilities. A primary motivation of our study is to explore differences (both in time and wavelength domains) among the observed and modeled SSI variability. By using differentials, i.e., irradiance changes, we lessen the need for an accurate reproduction of the `baseline' quiet-Sun spectrum.  Following \citet{liv2007}, one may assume that the `basal quiet atmosphere' is not measurably changed in the course of the solar cycle. 

The purely empirical models are frequently criticized for their simple approach of reducing the number of active components to two, namely the 
bright faculae/plages and the dark solar spots. While the sunspot component is determined from direct observations of sunspot areas and locations, the facular index is not. Rather the NRLSSI2-type models use an irradiance (i.e., disk-integrated) facular index, and assume that the changes in both active region and network faculae that affect solar irradiance similarly alter the (global) facular index. The NRLSSI2 model further assumes that once the relationship between solar irradiance variations and the sunspot and facular indices is determined at a given wavelength during solar rotation, the solar cycle changes in the indices then enable estimates of the solar cycle irradiance change at that wavelength. This use of the scaling factors derived from the rotational SSI variability for the long-term predictions is questioned as well. Below we show that such extrapolations are, at least at the present time, justifiable because the accuracy and long-term repeatability of the currently available SSI observations is inadequate for determining true solar cycle irradiance changes at most wavelengths. 

In regard to the `binary' approach of the purely empirical model, attempts to add more active components, in addition to, or different from, the active regions and distributed network emission that the global facular index encapsulates, face counter-productive challenges.  The SSI changes related to active network may be partially absorbed  by increases in the `effective' plage/facula area. On the other hand, contribution from the quiet network dominated  the Sun$^\prime$s magnetic flux in cycle 23 \citep{jin2011} and  modulated the long-term TSI changes by $\sim 30-40\%$ \citep{erm2003}.  Furthermore, the solar-cycle changes in the quiet network \citep{{erm2003},{sin2012}} are very hard to quantify due to the network$^\prime$s sizes and very low brightness contrasts. This may lead to the factor-of two differences in the estimates of the quiet-network filling factors \citep{{fou2001},{fou2011},{sin2012}}. The potential importance of this component, and the extent to which the adopted facular index does, or does not, include it, is yet to be fully and self-consistently addressed in terms of the models discussed in this paper; however note the alternative approach in \citet{fon2006} that, besides the active-Sun features (sunspot umbrae and penumbrae, faculae, plages) also employs various quiet-Sun categories: ``quiet-Sun cell interior, quiet-Sun network, active network'', following the definitions given in  \citet{fon2006}.   

The semi-empirical models employ detailed, quantitative assessment of the solar disk-projected areas occupied by various types of solar features, mainly judging the outcome by the feature$^\prime$s contrast, location and size as detected in solar imagery such as magnetograms. The brightness contrast of a particular region is governed by characteristic sizes of the contributing magnetic flux tubes \citep{{sol1993},{sol2013b}}. The contrast is usually treated as constant for a given class of features. However, there is a growing evidence that the contrasts depend on the feature$^\prime$s size, the wavelength of observation, the viewing angle, as well as the particular phase of the solar cycle \citep{{erm2003},{fou2011},{yeo2013},{yeo2014}}. Moreover, estimates of the filling factors are critically dependent on the instrument$^\prime$s spatial resolution \citep{cha2011}, requiring sub-arcsecond imaging of faint active regions and quiet networks,  thus rendering   some of the early-epoch observations of very limited practical use. One impediment for these types of models is the difficulty in achieving   long-term calibration stability of the solar images necessary to retrieve and quantify irradiance features over the duration of the solar cycle.

Evident from inspection of Figures 1 and 2, in the $\lambda >$350 nm region are OMI SSI changes that are systematically higher than the model estimates. This suggests that the OMI irradiances may include residual trends arising from the applied OMI degradation model, resulting in the observed 0.1-0.2\% long-term 
(on a $\sim$5 year time span) biases. We note the $\sim$0.1\% difference around 355-360 nm (Figure 1) between the outputs from the two partially overlapping OMI channels, UV2 and VIS, as well as the progressively growing in time deviations between the models and OMI observations  for $\lambda >$400 nm in Figure 2. The degradation model treats each OMI channel separately, thus producing the $\sim$0.1-0.2\% steps at the channel borders. The systematic 0.2-0.3\% model-observation deviations around 290-305 nm are likely caused by the known anomaly in the OMI wavelength registration \citep{{mar2014},{schen2016}}. 

But also evident in Figures 1 and 2 are noticeable differences between the OMI observations and the NRLSSI2 and SATIRE-S irradiance specifications in  wavelength regions that sample strong lines and line blends: e.g., Mg II and Mg I at 280, 285 nm, as well as multiple transitions in the 240-255 nm and 350-390 nm ranges. \citet{dan2016} remark on the SATIRE tendency to over-estimate the solar-cycle variability in strong UV spectral lines. Excluding these strong line blends and the regions with known instrument artifacts, we find an excellent (to within 1$\sigma$) agreement between the models and OMI observations, especially at the wavelengths sampled by the OMI UV2 channel, 310-360 nm.    

\begin{figure}
   \centering
    \includegraphics[width=1.0\columnwidth]{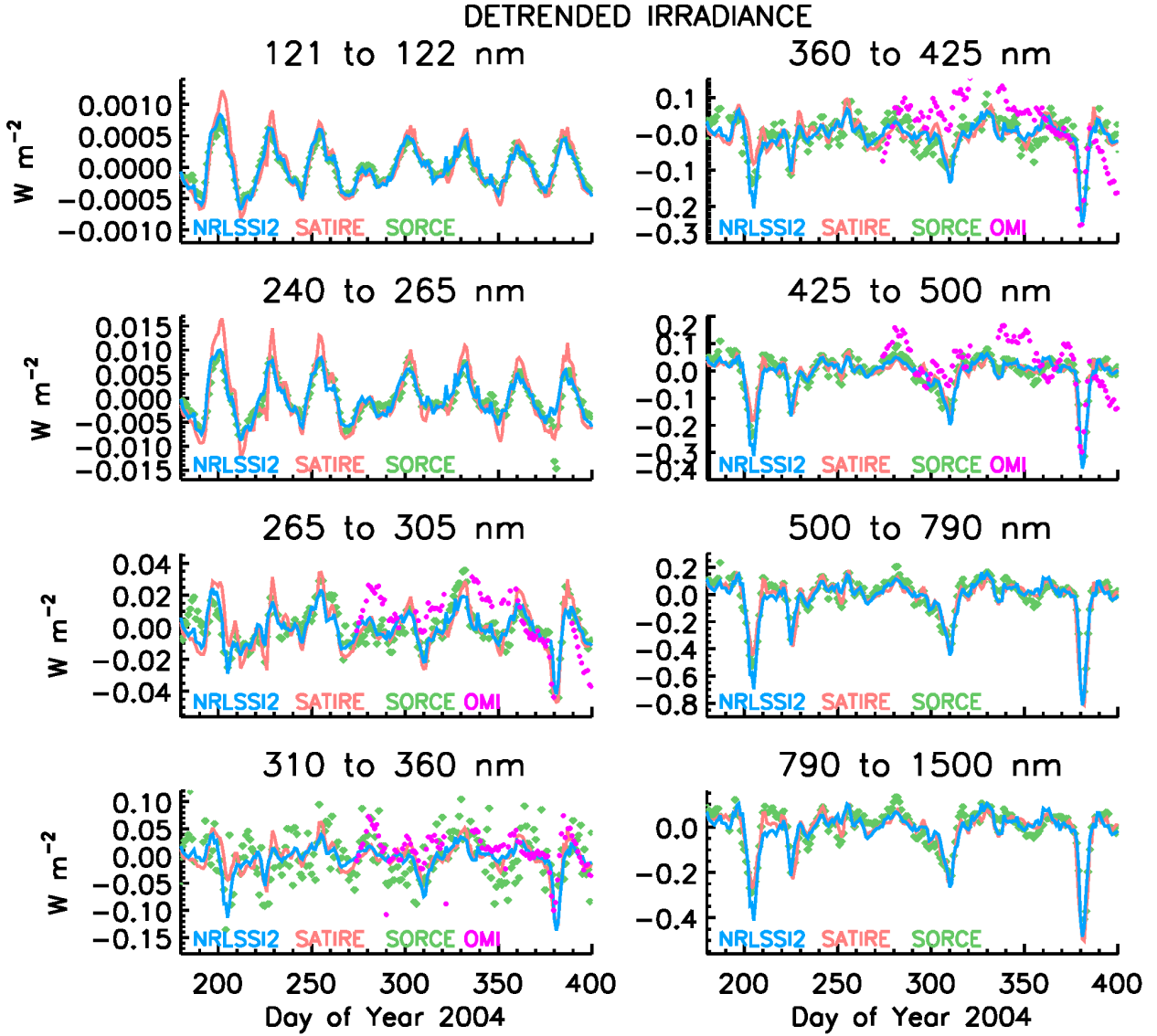}
   \caption{\small The de-trended, wavelength-binned SORCE and OMI data and consistently processed outputs from models, depicting the egress from the Cycle 23.} 
   \label{fig:fig4}
   \end{figure} 
The presented comparisons (cf. the wavelength-binned SSI changes in the $\lambda=$265-345 nm range in Figure 2) suggest that the NRLSSI2 model may underestimate, if slightly, the magnitude of SSI solar cycle variability in some UV spectral regions. Although independent studies \citep{yeo2015} have suggested that this was the case for the original NRLSSI model, the variability in the spectral region 300 to 400 nm is larger in the newly constructed NRLSSI2 than in NRLSSI. Nevertheless our analysis of OMI observations do imply larger solar cycle changes in this MUV region than in NRLSSI2, with the caveat that utilizing the OMI observations to specify solar cycle changes and assess differences between the modeled SSI changes during the solar cycle is more challenging than during solar rotation because of the possibility of unaccounted instrumental drifts. If real, such differences between OMI observations and the NRLSSI2 model may point to limitations in the model$^\prime$s approach of scaling wavelength-dependent rotational modulation to the solar cycle. Alternatively, the OMI dataset may overestimate (by $\sim$0.1-0.2\%) solar cycle changes in some spectral regions.  We again caution that these relatively small systematic errors and seeming synergy of the observed trends and model predictions should be considered as indicative rather than definitive, since the expected (at the $\sim$1 nm resolution) solar-cycle SSI variability falls well below the currently achievable $\sim$0.2\% accuracy limit in the $\lambda >$400 nm domain. Nevertheless, we regard this as a substantial improvement over the previously established $\pm$1\% limit \citep{{woods1996},{rott2004}} on identifying long-term solar changes.

\subsection{Short-Term (Rotational) SSI Variability: Observations and Models}

First of all, we note the almost perfect agreement (predominantly, to well within 1$\sigma$ errors) between the GOME-2 and OMI data depicting the short-term (rotational) SSI variability (Figure 3). This gives confidence that the observations provide reliable determinations of the magnitude of this variability. The observations and models also agree very well, to better than $\sim$0.05\%, in the $\lambda >$ 430 nm domain, as well as in the regions relatively devoid (e.g., 290-305 nm) of prominent spectral lines. Nevertheless, real differences are apparent among the observation and models. In the UV spectrum where the observed solar irradiance changes are largest, and the SSI variability measurements are thus relatively more precise, the SATIRE-S model systematically overestimates SSI changes during rotational modulation, relative to both observations and the NRLSSI2 model, with a notable exception of the underestimated Mg II doublet at 280 nm. SATIRE$^\prime$s tendency to overestimate the short-term changes in strong UV transitions was already noted by \citet{unruh2008}, suggesting that unaccounted non-LTE effects may play some role. Figures 4 and 5 further demonstrate that these differences persist in both solar cycles 23 and 24.  

The same result, though by a smaller margin, applies to the 
370-390 nm band (again, cf. Figure 3). In the strong Ca II lines (390-400 nm) neither model matches the rotational patterns observed by GOME-2 and OMI, even though the models and observations closely agree on the respective long-term changes. The observation-model differences at longer-wavelength regions are more subtle and more difficult to quantify, since they are frequently masked by gradually increasing (both in the wavelength and time domains) instrumental noise. 

Though here we sample relatively short periods from Cycles 23 and 24, we note that there is overall very good agreement between the rotational SSI changes observed at different solar cycles \citep{{del2012},{mar2014}}. Hence, these tendencies of the models relative to the observations are likely extendable on other epochs. For example, on the rotational timescales (three solar rotations in 2005) \citet{sol2013b} noted good agreement between the SORCE/SOLSTICE, SORCE/SIM and SATIRE-S SSI changes in the $\lambda >$300 nm domain. \citet{del2008} used scaling factors derived from rotational modulation variations during Cycles 21-22 to estimate long-term SSI behavior for comparison to their composite SSI data set \citep{del1998}.  While the composite SSI product does contain some artifacts \citep{pag2011}, the analysis presented in \citet{del2008} shows consistent behavior between observed long-term irradiance variations and proxy model predictions throughout Cycles 21-23, using observations from multiple instruments.  Similar results are shown for SSI measurements acquired at various solar cycles by \citet{del1998} and \citet{del2004}. Here we re-iterate the conclusion reached in \citet{mar2014}: the observed solar-cycle 24 and the appropriately scaled rotational SSI changes agree (excluding the few strong spectral features - see below) to within the provided uncertainties. We extend this conclusion on the rotational SORCE $\lambda>$500 nm data (Figure 3), as well as on the NRLSSI2 predictions for most of the sampled wavelengths.

   \begin{figure}
   \centering
    \includegraphics[width=1.0\columnwidth]{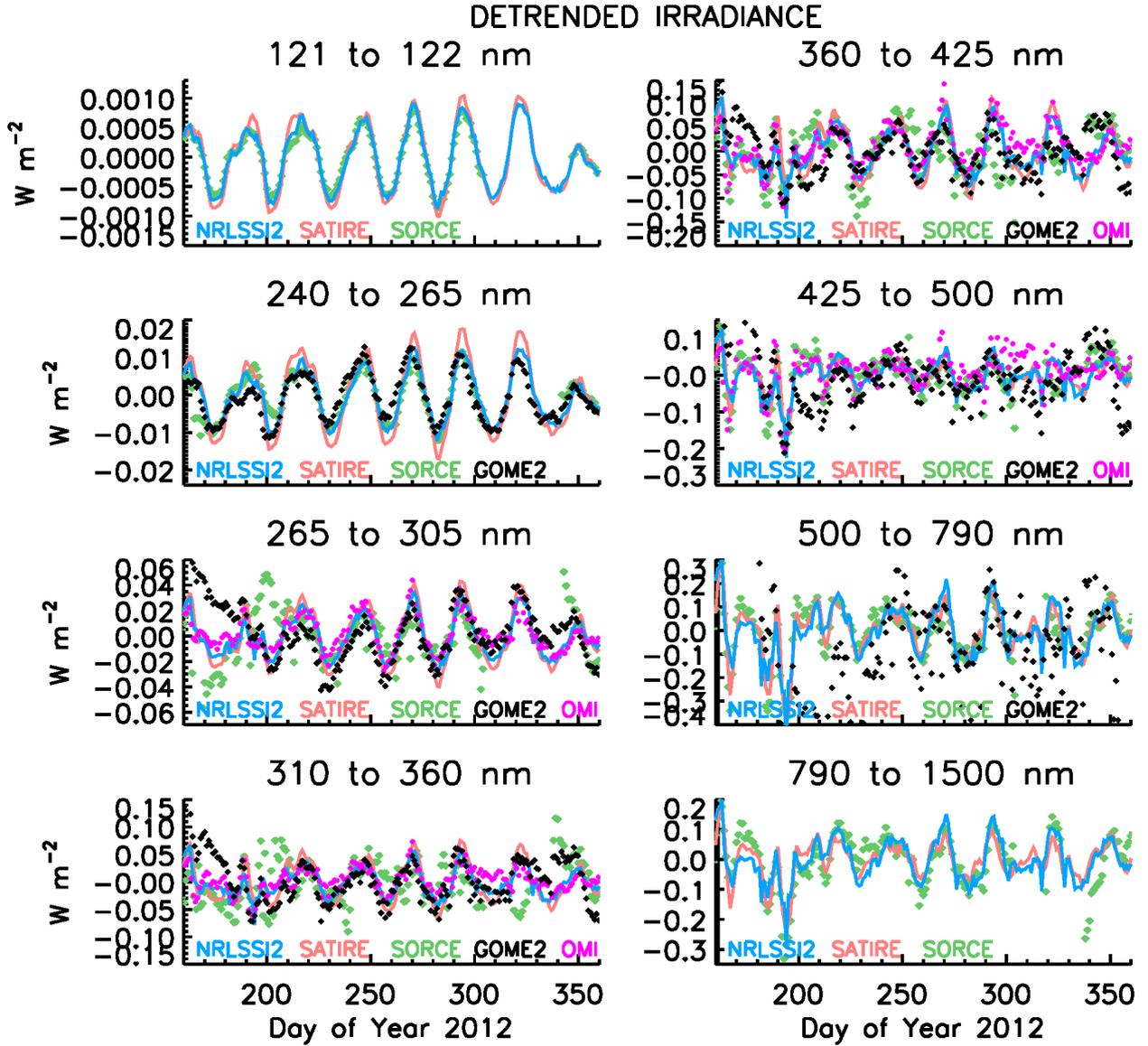}
   \caption{\small The same analysis as in Figure 4, now highlighting the rotational modulation patterns observed around the maximum of the Cycle 24 and including the de-trended GOME-2 data.} 
   \label{fig:fig5}
   \end{figure}

\subsection{Comparison of Short-Term and Long-Term Variability: OMI Observations and Models}

Current understanding of the sources of solar spectral irradiance variability suggests that the wavelength-dependence of changes during the solar cycle closely mimic the spectral dependence of the changes during solar rotation \citep{{flo2002},{rott2004},{del2004}}. This is because the changes on both time scales arise from changes in the amount of dark sunspots and bright faculae on the Sun. In the case of the solar cycle, the Sun$^\prime$s sub-surface dynamo produces different amounts of these features whereas the Sun$^\prime$s rotation imposes additional short-term modulation by altering the subset of the total population of these features projected to Earth. This assumption underlies the formulation of the NRLSSI2 model in which rotationally-modulated SSI changes are scaled to larger solar-cycle changes, according to sunspot and facular proxies. It also underlies the formulation of the SATIRE-S model in which the sunspot and facular contrasts remain constant, and irradiance variability accrues from the changing amount of bright and dark features on the Sun$^\prime$s disk. However, \citet{yeo2014} questions the validity of linear scaling of the rotation-induced changes observed in the chromospherically-sensitive lines (such as Mg II and H and K Ca II) to the SSI changes on solar-cycle timescales. 

\begin{figure}
   \centering
    \includegraphics[width=0.75\columnwidth]{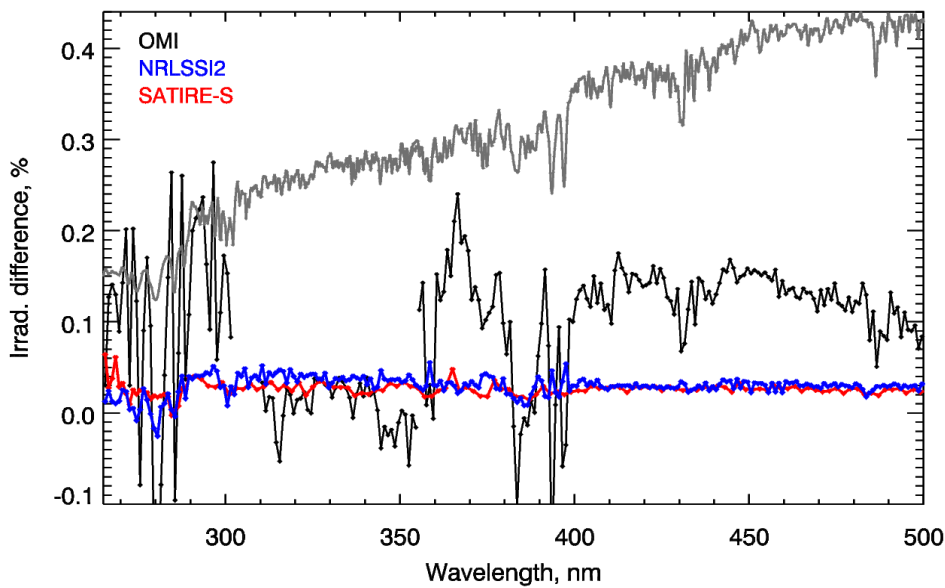}
   \caption{\small Differences between the long-term (yy2012-2014 vs. yy2007-2009) and short-term (8 rotational cycles in yy2012-2013) SSI changes for two models and OMI observations. An arbitrarily scaled solar spectrum (grey line) is shown for reference.} 
   \label{fig:fig6}
   \end{figure}

If present at all, differences in the short-term and long-term SSI changes have not yet been conclusively detected in observations, being frequently masked by uncertainties in measurements made by various instruments \citep{del2012}. Initial analysis of the more homogeneous OMI dataset \citep{mar2014} does suggest subtle deviations between long-term and short-term SSI patterns in the $\lambda > 350$ nm region. In the OMI data these deviations emerge when the SSI changes in spectral regions occupied by strong spectral features (mostly, line blends) are compared to changes in the adjacent relatively line-free regions. The rotational modulation tends to provide higher line/`continuum' contrasts compared to the solar cycle SSI modulation: e.g., for the CaII 393nm line the rotational line contrast reaches 0.85\%+/-0.15\% while the solar-cycle line contrast goes up to 0.60\%+/-0.15\%  \citep{mar2014}. Similar relative differences are observed in all strong spectral features in the $\lambda > 350$ nm range, thus alleviating significance of these subtle deviations. We expect that such line-contrast
changes would be more obvious in higher-resolution spectra: cf. the GOME-2 and OMI SSI line contrasts in Figure 5 from \citet{mar2014}. 

Here we ask how the rotational and solar-cycle SSI variability patterns compare in the NRLSSI2 and SATIRE simulations. 
To investigate this, we compare the short term and long-term model predictions (Figure 6) by equalizing  the corresponding model outputs (as seen in Figures 1 and 3) at a single wavelength point, chosen to be the top of the Mg I 285 nm line. I.e., shown in Figure 6 is the difference between (a) the long-term SSI changes from Figure 1 and (b) the short-term variability patterns (Figure 3) scaled by a single value. Such seemingly arbitrary scaling is justified by our goal of detecting any wavelength-dependent differences between the short-term and long-term variability patterns. The results do not change if we use Mg II 280 nm instead of the Mg I 285 nm line. 
  
Comparing the normalized rotational and long-term model output for each of the two models (Figure 6), we find that both models show small, $<0.05\%$, but systematic and very consistent deviations between the solar-cycle and rotational SSI variability. The magnitudes of these relative long-term vs. short-term SSI changes are practically wavelength-independent (save for the mild line effects in the $\lambda<$300 nm range in the NRLSSI2 data) and very similar for both models, despite the profound differences in the model assumptions and approaches. These small, but persistent long-term vs. short-term SSI differences may reflect changes in relative contributions of sunspots and faculae to solar rotational versus solar cycle variability. During solar rotational modulation, irradiance reductions associated with sunspots often exceed the irradiance enhancements associated with faculae; over the solar cycle the opposite is true. In the OMI spectra inter-channel (UV1, UV2 and VIS) instrumental biases obviously affect the long-term vs. short-term differences: note the $\sim$0.15\% steps at $\lambda\approx$300 nm and $\lambda\approx$360 nm. Besides, in line with the earlier reported trends in the OMI data [cf. Figures 3 and 6 from \citet{mar2014}], the observed long-term SSI changes show slightly lower line contrasts compared to the rotational SSI patterns: e.g., the strong line blends between $\lambda\lambda 380-400$ nm. These line-contrast features are not fully reproduced by the models.

\section{Conclusions}

Inter-comparing the observed (OMI, GOME-2 and SORCE) and modeled (NRLSSI2 and SATIRE-S) long-term (solar cycle) and short-term (solar rotational modulation) SSI changes, we conclude that overall, there is a fair- (at $\sim 2\sigma$ in some major absorption lines and blends) to-excellent (down to $1\sigma$, in relatively line-free spectra regions) agreement between the OMI observations and the models (NRLSSI2 and SATIRE-S) representing the SSI changes in the Cycle 24. The same excellent-to-fair agreement is also true for model-observation comparisons of the short-term (rotational) SSI changes, using the OMI, GOME-2 and SORCE data.  

In accord with \citet{yeo2015}, who found that NRLSSI under-estimates solar cycle changes, our results suggest that NRLSSI2 may slightly underestimate cycle 24 changes in the relatively line-free areas of the 265-290 nm region (Figure 1) when compared to the OMI data and the SATIRE-S predictions. However, the model (both NRLSSI2 and SATIRE-S) vs. observation (OMI) agreement improves to $<1\sigma$ (overall, better than 0.2\%) in the $\lambda>$300 nm domain (Figure 1).

On the solar-cycle timescale both models and OMI observations consistently demonstrate statistically significant, in-phase (all-positive) SSI variability in the $\lambda$310-380 nm range, contrary to the SORCE results (Figure 1). However, we emphasize that the longer-wavelength, $\lambda>$400 nm, OMI data and model predictions may be considered as indicative of in-phase variability only to within the quoted $\sim$0.2\% uncertainties (Figure 1). The same (in-phase SSI variability) applies to the short-term SSI changes. However, now SORCE data closely (to 0.05\%-0.2\%) follow the OMI and GOME-2 observations, once we take into consideration the large differences in spectral resolution (Figure 3).
 
We find that in both models the long-term (solar cycle) and appropriately scaled short-term (rotational) SSI variability patterns agree to better than 0.05\% in the $\lambda$265-500 nm range (Figure 6). This is rather surprising, considering the profound differences in the model approaches but may simply reflect different relative contributions of sunspots and faculae to solar irradiance variability on rotational versus solar cycle time scales.  

In complement to (and, sometimes, in replacement of) the relatively uncertain ($^<_\sim$0.2\%) solar-cycle data, the adequately accurate ($\sim$0.05\%), `de-trended' short-term observations [Figures 3-5; see also the composite $\lambda$170-795 nm SSI data from \citet{mar2014}] could be used in model verification. This, however, does not alleviate the urgent need for radically improved (thus at least matching the rotational accuracy) assessment of the solar-cycle SSI changes, especially in the $\lambda>$300 nm domain. 
 
Among other criteria, we suggest that the performance of SSI variability models should be judged by their ability to predict solar-cycle changes in strong UV transitions, since these are responsible for a significant proportion of the long-term variability of the total solar irradiance \citep{pre2002}. Indeed, the UV solar spectrum shows extensive line blanketing \citep{mitch1991}, and it is speculated that up to $\sim$60\% of the total irradiance variability is generated at $\lambda<$400 nm \citep{kriv2006}. The short-term, rotational SSI variability patterns seen in the OMI, GOME-2 and SORCE/SOLSTICE spectra can (and should) be used for further model improvements. 

Supporting the conclusions of \citet{unruh2008}, we show that, with few exceptions, for the rotational SSI changes SATIRE-S tends to over-estimate the variability in the strongest spectral UV lines and blends (cf. the $\lambda$300-450 nm range in Figure 3).

\begin{acknowledgements}
   We thank the anonymous referees for numerous helpful comments. Part of this work was supported by NASA contract NNG12HP08C. J. Lean acknowledges the support of the NASA SIST and NOAA CDR Programs, and appreciates ongoing collaboration with Odele Coddington in producing the NRLSSI2 CDR. We gratefully acknowledge Kok Leng Yeo for discussions and interpretation of the SATIRE-S model results.  The article used the SATIRE-S data available at https://www2.mps.mpg.de/projects/sun-climate/data.html. We also used the Aura/OMI data distributed via http://disc.sci.gsfc.nasa.gov/Aura, as well as the SSI measurements obtained by SORCE/SIM, SORCE/SOLSTICE (NASA) and EUMETSAT/GOME-2 (ESA). The editor thanks two anonymous referees for their assistance in evaluating this paper. 
\end{acknowledgements}


\end{document}